\documentstyle[preprint,aps,epsfig,tighten]{revtex}

\begin{document}

\title{Nature of the light scalar mesons}

\author{J. Vijande$^{1}$, A. Valcarce$^{1}$, F. Fern\'{a}ndez$^{1}$,
and B. Silvestre-Brac $^{2}$}
\address{$^{1}$ Grupo de F\'{\i}sica Nuclear and IUFFyM,
Universidad de Salamanca, E-37008 Salamanca, Spain}
\address{$^{2}$ Laboratoire de Physique Subatomique et de Cosmologie,
53 Avenue des Martyrs, F-38026 Grenoble Cedex, France}
\maketitle

\begin{abstract}
Despite the apparent simplicity of meson spectroscopy, light scalar
mesons cannot be accommodated in the usual $q\bar q$ structure.
We study the description of the scalar mesons below 2 GeV in terms
of the mixing of a chiral nonet of tetraquarks with 
conventional $q\bar q$ states. A strong 
diquark-antidiquark component is found for several states.
The consideration of a glueball as dictated by quenched 
lattice QCD drives a coherent picture of the isoscalar mesons.
\end{abstract}

\vspace*{2cm}
\noindent Pacs: 12.39.-x,12.38.-t

\maketitle

\newpage

\section{Introduction}

Nearly all known mesons made of $u$, $d$, and $s$ quarks fit neatly
into the multiplets expected in generic constituent quark models. 
The single striking exception are the scalar mesons, i.e., $J^{PC}=0^{++}$.
Either they form an anomalously light nonet (their masses do not fit
into the quark model predictions in its many variations) or a nonet and more 
scalar mesons appear in the energy region 1.2$-$1.5 GeV,
they overpopulate the expected number of states \cite{Jaf04}. 
The understanding of mesons with vacuum quantum numbers is a crucial
problem in low-energy QCD since they can shed some light on
the symmetry breaking mechanism in QCD and presumably also on confinement.

There are too many 0$^{++}$ mesons observed 
below 2 GeV to be explained as $q\bar q$ states.
There have been reported by the Particle Data Group (PDG) \cite{Pdg04}
two isovectors: $a_{0}(980)$ and $a_{0}(1450)$; five isoscalars:
$f_{0}(600), f_{0}(980), f_{0}(1370), f_{0}(1500)$ and $f_{0}(1710)$;
and three $I=1/2$ states: $K_{0}^*(1430)$, $K_{0}^*(1950)$ and recently
$K_0^*(800)$. The latest K-matrix fit of the $p \bar p$ 
annihilation data of Crystal Barrel Collaboration \cite{Ani03} gave 
rather definite information on the isoscalar resonances $f_0(600)$, $f_0(980)$, 
$f_0(1370)$\footnote{Quoted as $f_0(1300)$ in the original work.}, 
$f_0(1500)$,
$f_0(1710)$\footnote{Quoted as $f_0(1750)$ in the original work.}, 
and a broad state $f_0(1200-1600)$. Such state is introduced to
explain a bump observed simultaneously 
in the channels $\pi \pi$, $K \overline K$, $\eta \eta$ and $\eta \eta'$.
Its large width, of the order of $1200\pm400$ MeV, 
does not allow to determine reliably its mass.
Finally, the recent analyses by BES of the 
$J/\Psi \to \phi \pi^+ \pi^-$ and 
$J/\Psi \to \phi K^+ K^-$ data \cite{Bes04}
require a state $f_0(1790)$ distinct from the $f_0(1710)$. 
The experimental situation of the scalar mesons is
resumed in Table \ref{t1}.  

In contrast to the vector and tensor mesons, scalar resonances are
difficult to resolve. Their large decay widths
cause a strong overlap between resonances and background, which is
also obscured by 
the presence of several decay channels open up within a short mass interval.
In addition, the $K \overline K$ and $\eta \eta$ thresholds produce
sharp cusps in the energy dependence of the resonant amplitude.
Furthermore, theoretically one expects non$- q \bar q$ scalar objects, like
glueballs and multiquark states in the mass range below 2 GeV.
On the one hand, multiquarks have been justified to coexist 
with $q\bar q$ states in the energy region
around 1 GeV because they can couple to $0^{++}$, avoiding 
penalty due to orbital excitation \cite{Jaf77}. On the other hand,
lattice QCD in the quenched approximation predicts the
existence of a scalar glueball with a mass around 1.6 GeV 
\cite{Bal93}. 

While the low-energy hadron phenomenology has been successfully
understood in terms of constituent quark models, the scalar mesons
are still puzzling. 
Without exhausting the list of problems on the
scalar sector we recall that the quark structure as probed by
the electromagnetic interaction is definitively not
consistent with a naive $q \bar q$ composition \cite{Clo02}.
The interest and rather complicated situation of the scalar mesons
claims for a comprehensive study where other possible 
components are considered.
States of the quark model are most easily identified
with the hadrons we observe experimentally when 
unquenching (virtual hadron loop contributions) is
not important. Thus, for example, $\phi$ is 
dominantly an $s \bar s$ state, what follows from its decay to
$K \overline K$. Only a small fraction of the Fock space
of the physical $\phi$ is $K \overline K$. This is in part
due to the $P-$wave nature of the $K \overline K$ component.
In contrast, scalar mesons are strongly affected by their
coupling to open hadron channels. The fact that the
$f_0(980)$ and the $a_0(980)$ couple to both 
$\pi \pi /\pi \eta$ and $K \overline K$ channels means that
virtual hadron loop contributions are important.
As a consequence, conventional $q \bar q$ states 
are expected to mix with four-quark ($qq\bar q \bar q$) 
states to yield physical mesons.
A color singlet four-quark state
can be obtained in two different coupling schemes, 
$\left[ (q q) (\bar q \bar q) \right]$ or
$\left[ (q \bar q) (q \bar q) \right]$. While in the first case
the singlet color states are obtained from the 
$ \left[ (q \otimes q)_6 (\bar q \otimes \bar q)_{\bar 6} \right]$ 
(the subindex standing for the color state) and
$ \left[ (q \otimes q)_{\bar 3} (\bar q \otimes \bar q)_{3} \right]$ 
couplings, with a
non-intuitive physical interpretation, in the second case the
total singlet color states are driven by
$\left[ (q \otimes \bar q)_1 (q \otimes \bar q)_{1} \right] \equiv (MM) $ and
$\left[ (q \otimes \bar q)_{8} (q \otimes \bar q)_{8} \right] \equiv (QQ) $, where 
the physical interpretation is made evident. The first 
component will correspond to a molecule of mesons, while the
second one is a compact component that is not factorizable into
singlet color meson-meson channels.  

In this work we will address the study of hadrons with 
zero baryon number described as clusters of constituent (massive) quarks 
confined by a realistic interaction. We make use of
a standard constituent quark model applied to the study
of the nonstrange baryon spectra and the baryon-baryon interaction \cite{Rep05}. 
This model has been recently
generalized to all flavor sectors giving a reasonable description
of the meson spectra except for the scalar mesons \cite{Vij03}.
The model is strongly constrained by its application to other hadron sectors, 
representing in this way an advance with respect to similar recent studies based on models 
explicitly designed to study the scalar mesons \cite{Nap04}. The paper is organized 
as follows. In the next
section we will resume the most relevant aspect of the constituent quark model
used. Sect. III will be devoted to present and discuss the results. Firstly we
will analyze the results based on a naive $q\bar q$ scheme, then we will
consider the mixing to four-quark components and finally we will include 
a scalar glueball. In Sect. IV we will conclude summarizing our main findings.

\section{SU(3) constituent quark model}

The model is based on the assumption
that the constituent quark mass appears because of the spontaneous
breaking of the original $SU(3)_{L}\otimes SU(3)_{R}$ chiral symmetry at
some momentum scale, which is
the most important nonperturbative phenomenon for hadron structure at low
energies. In this domain of momenta quarks are quasiparticles with a constituent
mass interacting through scalar and pseudoscalar boson-exchange potentials.

Beyond the chiral symmetry breaking scale one expects the dynamics
being governed by QCD perturbative effects. They are taken into account
through the one-gluon-exchange (OGE) potential. 
Following de R\'{u}jula {\it et al.} \cite{ruju}
the OGE is a standard color Fermi-Breit interaction.

Finally, any model imitating QCD should incorporate confinement. 
Lattice calculations in the quenched
approximation derived, for heavy quarks, a confining interaction linearly
dependent on the interquark distance. The consideration of sea quarks apart
from valence quarks (unquenched approximation) suggest a screening effect on
the potential when increasing the interquark distance \cite{Bal01}. Creation of light-quark
pairs out of vacuum in between the quarks becomes energetically preferable
resulting in a complete screening of quark color charges at large distances.
Although string breaking has not been definitively confirmed
through lattice calculations \cite{bali2}, a quite rapid
crossover from a linear rising to a flat potential is well
established in SU(2) Yang-Mills theories \cite{este}. Such screened 
confining potentials provide with an explanation to the
missing state problem in the baryon spectra \cite{miss}, improve the
description of the heavy meson spectra \cite{Vij04}, and justify the 
deviation of the meson Regge trajectories from the linear behavior
for higher angular momentum states \cite{golr}.

Explicit expressions of the interacting potential derived from the
nonrelativistic reduction of the lagrangian on the static approximation
and a more detailed discussion of the model can be found in Ref. \cite{Vij03}.

\section{Results}

In nonrelativistic quark models, gluon degrees 
of freedom are frozen and therefore the wave 
function of a zero baryon number (B=0) hadron may be written as
\begin{equation}
\label{mes-w}
\left|\rm{B=0}\right>=
\Omega_1\left|q\bar q\right>+\Omega_2\left|qq\bar q \bar q\right>+....
\end{equation}
where $q$ stands for quark degrees of freedom and, as mentioned 
above, the coefficients $\Omega_i$ will take into account the possibility
that four-quark states could mix with $q \bar q$ states in the 1 GeV energy
region. $\left|\rm{B=0}\right>$ systems could then be 
described in terms of a hamiltonian 
\begin{equation}
H=H_0 + H_1 \,\,\,\,\, {\rm being} \,\,\,\,\,  
H_0 = \left( \matrix{H_{q\bar q} & 0 \cr
0 & H_{qq\bar q\bar q} \cr } \right) \,\,\,\,
H_1 = \left( \matrix{0 & V_{q\bar q \leftrightarrow qq\bar q\bar q} \cr 
V_{q\bar q \leftrightarrow qq\bar q\bar q} & 0 \cr } \right)
\label{eq1}
\end{equation}
where, as will be shown below, 
the non-diagonal terms can be treated perturbatively, 
allowing to solve therefore the two- and four-body sectors separately.

\subsection{The $q\bar q$ sector}

For the two-body problem, we have solved the Schr\"odinger equation 
using the Numerov algorithm \cite{Num00}.
As mentioned above the model provides with a correct description of the 
full meson spectra except for the scalar states \cite{Vij03}. 
In Table \ref{t2} we compare the masses of the scalar
$q \bar q$ states to experimental data.
In the last two columns we indicate the name and the mass
of the hypothetical experimental state corresponding to the theoretical result.

Let us examine a possible correspondence between $q \bar q$ states
and $0^{++}$ experimental states. The isovector sector is spectroscopically
simple. The $a_0(980)$ would correspond to the $^3P_0$ member of the $1\,\,^3P_J$
isovector multiplet. The $a_{0}(1450)$ should be considered as the scalar
member of the $2\,^3P_J$ excited isovector multiplet. 
This reinforces the prediction of the quark model, the spin-orbit
force making lighter the $J=0$ states with respect to the $J=2$.
The assignment of the $a_0(1450)$ to the scalar member of the 
$1\,\,^3P_J$ multiplet \cite{Clo02,Gei91}
would contradict this idea, because the $a_2(1320)$ is a well
established $q \bar q$ pair and much lower in energy than the
$a_0(1450)$. The same behavior is evident in the
$c\bar c$ and $b\bar b$ spectra, making impossible to
describe the $a_0(1450)$ as a member of the $1\,\,^3P_J$
isovector multiplet without spoiling the description 
of heavy-quark multiplets \cite{Vij03}.
However, there appear several problems with the decay patterns. For
example, the $a_0(980)$, considered as a pure light $q\bar q$ state,
is known to underestimate the $\phi \to \gamma a_0(980)$ width by
one or two orders of magnitude \cite{Bar02}.

Concerning the $I=1/2$ sector, as a consequence of the larger mass
of the strange quark, the quark model predicts a mass for the 
lowest $0^{++}$ state 200 MeV greater than the
$a_{0}(980)$ mass. Therefore, being the $a_0(980)$ the
member of the lowest isovector scalar multiplet, the $K_0^*(800)$
cannot be explained  as a $q\bar q$ pair.

The most obvious problems appear in the case of the isoscalar states.
It is clear from Table \ref{t2} the absence of theoretical isoscalar
$q \bar q$ states in three different energy regions. No states
are predicted between 0.5$-$1.3 GeV, where the $f_0(980)$
resides, neither between 1.4$-$1.7 GeV, the energy range 
of the $f_0(1500)$, nor between 1.9$-$2.2 GeV, 
where the $f_0(2100)$ is placed. 

Similar conclusions concerning the $f_0(980)$ 
and the $K_0^*(800)$ have been obtained using the extended
Nambu-Jona-Lasinio model in an improved ladder approximation of the
Bethe-Salpeter equation \cite{Ume03}. This indicates that relativistic
corrections may not improve the situation.
Therefore, a different structure rather than a naive 
$q\bar q$ pair seems to be needed. In particular, the $f_0(980)$
has been suggested as a possible four-quark state \cite{Ach00}.

\subsection{Including four-quark states}

Let us examine if the naive $q\bar q$ picture supplemented by
four-quark states could acknowledge for the experimental
situation. The four-body problem has been solved by means of 
a variational method using as trial wave function 
the most general linear combination of gaussians \cite{Suz98}. 
In particular, the so-called ``mixed terms'' (mixing the various
Jacobi coordinates) that are known to have a great influence in
the light quark case have been considered. 
The method to solve the
four-body problem has been tested in the case of a system
whose quantum numbers can only be obtained by means 
of a multiquark state, the isospin two $X(1600)$. This state has been 
observed in the reaction $\gamma\gamma\rightarrow\rho\rho$ near
threshold, reported with a mass of $1600\pm100$ MeV and
quantum numbers $I^G(J^{PC})=2^+(2^{++})$ \cite{Pdg04}. 
It cannot be described as a $q\bar q$ state, being 
therefore an exotic meson that can be understood as 
four light quarks coupled to $I=2$, $S=2$ and $L=0$. 
The energy obtained for this configuration is 1500 MeV,
in agreement with the experimental data, giving 
confidence to the results obtained in the four-quark calculation.

The masses and flavor dominant component obtained for the
low-lying four-quark states 
are shown in Table \ref{t2}, compared to the possible
experimental assignment. As can be seen, for the isoscalar
mesons the four-quark states appear precisely in the energy 
region where no $q \bar q$ states are predicted.
In spite of that, neither the $q \bar q$ nor the $qq\bar q \bar q$ configurations 
match the experimental data. In particular, the $f_0(980)$ cannot be understood
as a $q \bar q$ pair without failing to describe the
isovector sector. Nonetheless, it is worth to notice that the 
nearly mass degeneracy observed experimentally between the $a_0(980)$
and the $f_0(980)$ could be explained spectroscopically if the
$a_0(980)$ is considered as a $q\bar q$ pair and the
$f_0(980)$ as a four-quark state, although such assignment
fails to explain the observed properties of these states.

The Hamiltonian $H_1$ in Eq. (\ref{eq1}) describes the
mixing between two- and
four-body configurations. Its explicit expression would require
the knowledge of the operator annihilating a quark-antiquark pair 
into the vacuum. This could be done, for example, using a $^3P_0$ model,
but the result will always depend on the parametrization used to describe
the vertex. For simplicity, we have parametrized this mixing 
by looking to the quark pair that it is annihilated, and
not to the spectator quarks that will form the final $q\bar q$ state: 
\begin{eqnarray}
\left<nn\bar n\bar n|V|n\bar n\right>=\left<ns\bar n\bar s|V|s\bar s\right>
=\left<nn\bar n\bar s|V|n\bar s\right>&=&C_n\\ \nonumber
\left<ss\bar s\bar s|V|s\bar s\right>=\left<ns\bar n\bar s|V|n\bar n\right>
=\left<ns\bar s\bar s|V|n\bar s\right>&=&C_s.
\end{eqnarray}
The mixing parameters, $C_s$ and $C_n$, are chosen to drive 
the $f_0(980)$ state to its physical mass. 
A fine tune of the $q \bar q$ model has been done to maintain the description of the
isovector sector. In particular, the mass obtained for the bare $1P$ $q \bar q$ 
state (that would represent the $a_0(980)$ as a two-quark state)
is 1079 MeV. The final resulting values are $C_n=165$ MeV and $C_s=70$ MeV. 
Taking into account the degeneracy observed in Table \ref{t2} between the
$a_0(980)$ and the $f_0(980)$, this correction is consistent with the assumption that 
$H_1$ can be treated perturbatively. 

The obtained mass and dominant flavor component for all 
the scalar mesons are given in Table \ref{t4}. The interpretation of the light
scalar mesons in terms of two- and four-quark components
allows for an almost one-to-one correspondence between theoretical
states and experiment. A dominant tetraquark component is found for 
the $f_0(980)$, $f_0(1500)$, $a_0(1450)$ and the $K_0^*(1430)$,
and an important tetraquark component is predicted for the $a_0(980)$,
although it has a dominant $q \bar q$ structure (see
Table \ref{t5}). The four-quark structure of the $a_0(980)$ and
$f_0(980)$ would avoid the underestimation of the partial width
for the decays $\phi \to \gamma f_0(980)$
and $\phi \to \gamma a_0(980)$ obtained in the case of a pure 
$q\bar q$ component \cite{Bar02}. In the case of the $f_0(980)$ it would also make it
compatible with the similar branching ratios observed
for the $J/\psi \to f_0(980)\phi $ and $J/\psi \to f_0(980)\omega$
decays \cite{Ach00}. 
Concerning the $f_{0}(1370)$ (which has already been suggested as 
corresponding to two different states \cite{Bla01}) we obtain two
states around this energy, the heavier one with a dominant nonstrange
content which favors its assignment to the $f_0(1370)$;
the other with a high $s\bar s$ content without experimental
partner in the PDG. This state, which couples strongly to the
$K\overline K$ channel, may correspond to the broad resonance 
$f_0(1200-1600)$ predicted by Anisovich and collaborators \cite{Ani03}.

Our results support the conclusion of the non-$q\bar q$
structure of the $f_0(1500)$ \cite{Ams02}. 
The $f_0(1500)$ has not been measured either in
$\gamma\gamma\to K_SK_S$ by L3 Collaboration \cite{Acc01} or in
$\gamma\gamma\to\pi^+\pi^-$ by ALEPH Collaboration \cite{Bar00}, implying
that this state, if quarkonium, should be dominantly $s\bar s$.
However, the branching ratios for the decay into two pseudoscalar mesons are
only compatible with an almost pure $n\bar n$ structure. One should notice that
the experimental situation for the two-photon decay is not
definitively settled, since no partial wave analysis has been performed either by L3 or
ALEPH and the possible coupling of $f_0(1500)$ and $f_0(1370)$ has not been
considered.
Nonetheless, there are other experimental evidences supporting the non-$q\bar q$ 
structure of the $f_0(1500)$.
For instance, in the central production experiments performed by the WA102
Collaboration \cite{Bar97} it has been discovered
a kinematic filter able to discriminate between
$P-$wave $q\bar q$ states and $S-$wave states like glueballs, $qq\bar q\bar
q$ multiquarks or $K\overline K$ molecules \cite{Clo97}. Its essence is
that the pattern of resonances depends on the vector difference of the transverse
momentum recoil of the final state protons ($dP_T$). When $dP_T$ is 
large $P-$wave $q\bar q$ states are prominent, whereas at small $dP_T$ they
are suppressed. Following this discovery there has
been an intensive experimental program that has reported that the only isoscalar scalar 
states that appear in the low $dP_T$ regime are the $f_0(980)$, $f_0(1500)$ and $f_0(1710)$, therefore
confirming their non$-q\bar q$ structure.
Another experimental evidence has also been obtained from the analysis of the data of 
the WA102 Collaboration \cite{Clo01}. 
The azimuthal dependences as a function of $J^{PC}$ and the momentum
transferred at the proton vertices appear to divide the scalar mesons into two
classes: the $f_0(980)$, $f_0(1500)$ and $f_0(1710)$ which are all strongly
peaked at small angles and the $f_0(1370)$ which peaks at large angles, what
may indicate a different nature of these states.

Finally, we obtain a dominant $n\bar n$ state corresponding to the $f_0(1790)$. 
This energy region is expected to allocate the radial excitation of the 
$n \bar n_{2P}$ state, $f_0(1370)$. We also find a candidate
for the $f_0(2200)$, experimentally identified as an
$s \bar s$ state \cite{Ani00}, with an energy of 2212 MeV.

A crucial test of the quark structure of the scalar mesons would be
the systematic study of two-photon decay widths of neutral scalars,
which is still lacking. The two-photon decay is dominated by the 
$q\bar q$ component of the wave function (the four-quark
component is known to give a small contribution \cite{Ach82}). 
In this case the scalar two-photon decay width 
can be related to the well-known experimental data
of the $J=2$ multiplet member through \cite{Tum53}
\begin{equation}
\label{eq4}
\Gamma_{\gamma\gamma}\left( 0^{++} \right) =
k \, \left( m_0 \over m_2 \right)^3
\Gamma_{\gamma\gamma}\left( 2^{++} \right) 
\end{equation}
with obvious notation.
The factor $k$ arises from spin multiplicities and the consideration
of relativistic corrections to $O(v^2/c^2)$ in three different approaches
leads to a value lower than the corresponding nonrelativistic one, $k\approx2$
as compared to $k=15/4$ \cite{Ber83}.
Data on the charmonium states $\chi_{c2}$ and $\chi_{c0}$ using the
nonrelativistic value are in good agreement with Eq. (\ref{eq4}).
Using the results given in Table \ref{t5} we obtain
\begin{equation}
\Gamma_{\gamma \gamma} \left[ a_0(980) \right] 
= 0.65 \pm 0.04 \,\, {\rm keV} \,\,\,\, ; \,\,\,\,
\Gamma_{\gamma \gamma} \left[ f_0(980) \right] 
= 0.23 \pm 0.02 \,\, {\rm keV} \,
\end{equation}
that compares rather well with the experiment \cite{Pdg04},
\begin{equation}
\Gamma_{\gamma \gamma}^{\rm Exp} \left[ a_0(980) \right] 
= 0.3 \pm 0.1 \,\, {\rm keV} \,\,\, ; \,\,\,
\Gamma_{\gamma \gamma}^{\rm Exp} \left[ f_0(980) \right] 
= 0.39^{+ 0.10}_{- 0.13} \,\, {\rm keV} \, .
\end{equation}
Let us note that the experimental width for the decay
$a_0(980) \to \gamma \gamma$ requires the knowledge of 
the branching ratio of the decay $a_0(980) \to  \eta \pi$, that
has been taken to be 0.24 $\pm$ 0.08 keV. Small variations on this
number will induce large modifications on the $a_0(980)$ 
two-photon decay width.
These results would be modified by the $K\overline{K}$ component of
the four-quark wave function \cite{Bar85}. However, this component is rather
small, less than 1 \%, what would not modify the results significantly.

The final physical picture arising shows 
an involved structure for the flavor wave function of the
light scalar mesons, in agreement with their complicated pattern
decays. In Table \ref{t5} we give a
detailed description of the flavor wave function of some
selected scalar states. Regarding the four-quark structure it
presents at the same time compact and molecule components.
For example the 72 \% of $(ns\bar n \bar s)$ of the $f_0(1500)$ 
can be decomposed in the following way
$f_0(1500) = 0.723 \left\vert MM \right\rangle 
+ 0.691 \left\vert QQ \right\rangle$, where the meson-meson
component involves the asymptotic states: $\eta' \eta'$, $\eta \eta$,
$K \overline K$, $\eta \eta'$, $\omega \phi$ 
and $K^* \overline K^*$, generating
decays that will feed a wide range of physical states. 
The meson-meson component contains a weakly bound
$K\overline K$ molecule which may decay by annihilation
through an intermediate $0^{++}$ $q\bar q$ meson 
into two pions \cite{Wei90} as observed experimentally.
The same reasoning could be applied to the $f_0(980)$ and $a_0(980)$ 
in such a way that they would be at the same time
four-quark states and $K \overline K$ molecules
as suggested by the nearness of the $K \overline K$
threshold \cite{Clo02}. Finally, let us stress
the presence of an important diquark-antidiquark,
$\left[ (q \otimes q)_{\bar 3} (\bar q \otimes \bar q)_{3} \right]$,
component in the four-quark wave function (57 \% for 
the case of the $f_0(1500)$ discussed above), where QCD
predicts a strong attraction in the $S-$wave 
when in a flavor nonet \cite{Jaf04,Jaf79}. 
The presence of a diquark-antidiquark component
in the four-quark wave function of the scalar mesons
has been recently recognized in a schematic calculation
of the scalar mesons \cite{Mai04}.

In the literature one finds alternative approaches to understand
the rather complicated scenario of the scalar mesons. An earlier
attempt to link the understanding of the $NN$ interactions with meson
spectroscopy was done based on the J\"ulich potential model \cite{PRD52}. The
structure of the scalar mesons $a_0(980)$ and $f_0(980)$ was investigated in
the framework of a meson exchange model for $\pi\pi$ and $\pi\eta$ scattering.
The $K\overline K$ interaction generated by the vector-meson exchange, which for
isospin $I=0$ is strong enough to generate a bound state is much weaker for
$I=1$, making a degeneracy of $a_0(980)$ and $f_0(980)$ impossible, as found
in our model as $q\bar q$ pairs. Although both scalar mesons result from the coupling to the $K\overline K$ 
channel explaining in a natural way their similar properties, the underlying
structure obtained was, however, quite different. Whereas the $f_0(980)$
appears to be a $K\overline K$ bound state the $a_0(980)$ was found to be a dynamically 
generated threshold effect. Similar conclusions have been obtained in a
chiral unitary coupled channel approach, where the $f_0(600)$, the $a_0(980)$, and the 
$K^*_0(800)$ rise up as dynamically generated resonances, being poles in the $S-$wave meson-meson 
scattering amplitudes, while the $f_0(980)$ is a combination of a strong
$S-$wave meson-meson unitarity effect and a preexisting singlet resonance \cite{Ose97}.
  
In a different fashion within the quark model the same problem was
illustrated in Ref.\cite{TORN}. The bare mass used for the $n\bar n$ pair is much larger
than the $a_0(980)$ and $f_0(980)$ experimental masses. It is the effect of
the two-pseudoscalar meson thresholds the responsible for the substantial shift
to a lower mass than what is naively expected from the $q\bar q$ component alone.
This gives rise to an important $K\overline K$ and $\pi\eta'$ components in the
$a_0(980)$ and $K\overline K$,  $\eta\eta$, $\eta'\eta'$ and $\eta\eta'$ in the
$f_0(980)$. In particular for the $a_0(980)$ they obtain the $K\overline K$
component to be dominant near the peak, being about $4-5$ times larger than
the $q\bar q$ component. A similar conclusion, that the description of the
$a_0(980)$ and $f_0(980)$ requires from more complex structures, is also
obtained from our analysis. 

\subsection{Including the light scalar glueball}

Due to its non-abelian character QCD predicts the existence of isoscalar
mesons containing only gluons, the glueballs $(G)$. Therefore one
should finally wonder whether there is some place for them in
the meson spectra. There are several alternatives in the literature concerning these states,
although there seems to exist a consensus that the lowest
glueball should be the scalar one \cite{Wes96}. Some of these approaches predict the existence of a
low-lying glueball below 1 GeV \cite{Ven04}, while others seem to prefer an scenario 
where the glueball appears in a higher energy region, around 1.4$-$1.8 GeV \cite{Bal01,Isg85,Mcn00,Lee00}. 
In our description of the scalar isoscalar states the second prescription is
preferred. As can be seen in Table \ref{t4}, there is only one experimental state
without theoretical partner, being its energy around 1.5 GeV.
Lattice QCD in the quenched approximation predicts the existence of a
scalar glueball with a mass around 1.6 GeV \cite{Bal01}. Besides, recent lattice 
studies confirm that there is indeed significant mixing between $G$ and 
the nearest $q \bar q$ states, together with associated
mass shifts in the $J^{PC}=0^{++}$ sector \cite{Mcn00}. 
For pure gluonium one expects coupling of similar strengths to $s\bar s$
and $n\bar n$ mesons since gluons are flavor blind. Lattice calculations derived
a $G - q\bar q$ mixing nearly flavor blind with a maximum deviation of
20\% \cite{Lee00} and suggested that the preferred glueball mass falls into the range
\begin{equation}
{{M_{s \bar s} + M_{n \bar n}} \over 2} > M_G > M_{n \bar n} \, ,
\label{glue}
\end{equation}
where $M_{n\bar n}$ and $M_{s \bar s}$ are the bare 
quarkonium masses (before mixing). 
All these results obtained from lattice QCD indicate than
glueballs and their flavor-mixing are a controlling
feature of the meson spectroscopy in the 1.3$-$1.7 GeV mass region.

Therefore, based on intuition from lattice QCD \cite{Bal01,Mcn00,Lee00,Ams95,Clo00}
we have investigated the mixing between a $J^{PC}=0^{++}$ glueball and the
$q \bar q$ nonet in its vicinity. We have included in our calculation
a scalar glueball satisfying the Eq. (\ref{glue})
and with the mixing to the $2P$ $n \bar n$ and $s \bar s$ pairs predicted by
lattice QCD \cite{Lee00}. In particular, within the model used
the bare $n \bar n$ state has a 
mass of 1503 MeV, and the bare $s \bar s$ state 1850 MeV, giving
a mass for the scalar glueball between 1503 and 1676.5 MeV. We have used
$M_G =$ 1643 MeV. Besides as suggested in Ref. \cite{Lee00} we take
$\left\langle G \right\vert V \left\vert n \bar n \right\rangle = 
\sqrt{2} \, r \, \left\langle G \right\vert V \left\vert s \bar s \right\rangle$ 
and $\left\langle G \right\vert V \left\vert s \bar s \right\rangle =$ 64 MeV
and $r \approx$ 1.2 ($r=$1 for exact $SU(3)$ symmetry, exact flavor blind).
The results are resumed in Table
\ref{t8} noting a correspondence between theoretical 
predictions and experiment. This assignment suggests that there are four
isoscalar mesons that are not dominantly
$q \bar q$ states, they are the 
$f_0(980)$ (dominantly a $nn \bar n \bar n$  state), the
$f_0(1500)$ (dominantly a $ns \bar n \bar s$  state), the
$f_0(1710)$ (dominantly a glueball) and the
$f_0(2020)$ (dominantly a $ss \bar s \bar s$  state). 
This is clearly seen in Fig. \ref{fig1} where we have
constructed the two Regge trajectories associated to 
the isoscalar mesons. As it is observed the masses of the
$f_0(600)$, $f_0(1200-1600)$, $f_0(1370)$, $f_0(1790)$, $f_0(2100)$, $f_0(2200)$ fit
nicely in one of the two Regge trajectories, while those
corresponding to the $f_0(980)$, $f_0(1500)$, $f_0(1710)$, $f_0(2020)$ do not
fit for any integer value. The exception would be the $f_0(2020)$
that it is the orthogonal state to the $f_0(2100)$ having
almost 50\% of four-quark component. 

The glueball component is shared between the three neighboring states: 20 \% for the $f_0(1370)$,
2 \% for the $f_0(1500)$ and 76 \% for the $f_0(1710)$.
Our results assigning the larger glueball component to the $f_0(1710)$ are on
the line with Refs. \cite{Mcn00,Lee00} and differ from those of
Refs. \cite{Ams02,Ams96} concluding that the $f_0(1710)$ is dominantly $s \bar s$.
One should notice that none of these studies consider the recent result of the
BES Collaboration suggesting the existence of a scalar state close to the
$f_0(1710)$ that could be the $q\bar q$ dominant one. It should be also
mentioned that most studies in the literature \cite{Ams02,Mcn00,Lee00,Ams96} do only address 
a particular set of scalar states that they consider could be identified with the glueball.
This makes clear the complicated situation in the scalar sector
with several alternative interpretations.

\section{Conclusions}

In summary, the spectroscopy and many of the decay properties
of the scalar mesons can be understood under the hypothesis of 
the mixing between two- and four-quark components. If one
considers the existence of a scalar glueball within the intuition
of lattice QCD in the quenched approximation, the description
is improved. Our results suggest the existence of the state
recently reported by BES Collaboration, $f_0(1790)$, as the radial excitation of the
$f_0(1370)$ and also show evidence for the presence of a new
scalar state around the $f_0(1370)$ already suggested by Ref.
\cite{Bla01} and definitively predicted by the K-matrix
analysis of the $p\bar p$ annihilation data of the
Crystal Barrel Collaboration, $f_0(1200-1600)$.
Our description of the $f_0(980)$, $f_0(1500)$ and $f_0(1710)$ as 
non-$q\bar q$ states is in agreement with several experimental evidences, 
in particular those obtained by the WA102 Collaboration.
The final flavor structure of the light scalar mesons becomes rather involved allowing
to deep the understanding of their complicated decay patterns.
A strong diquark-antidiquark component is found.
Some problems remain open as it could be the high strange content
predicted for the $f_0(2020)$.
The study of radiative transitions and two-photon decay widths
should help to clarify the flavor mixing 
and the nature of the $I=0$ scalar sector. 

Our work presents an interpretation
of the scalar mesons in a model constrained by the description of other hadron
sectors. It drives to a final scenario that it is compatible with some
other models in the literature and it differs from the results of others.
The final answer could only be obtained from precise experimental data that would 
allow to discriminate between the predictions of different theoretical models. 
The set of data is so huge, and sometimes so poor, that one always may find a positive or
negative interpretation of some of them. 


\section{acknowledgments}
After this work was completed we learned that similar ideas regarding the $f_0(1790)$ have been recently suggested by 
F.E. Close and Q. Zhao \cite{Clo05}.
This work has been partially funded by
Ministerio de Ciencia y Tecnolog{\'\i}a
under Contract No. FPA2004-05616, by Junta de
Castilla y Le\'{o}n under Contract No. SA104/04,
and by a IN2P3-CICYT agreement.

\begin{table}[tbp]
\caption{Experimentally reported light scalar mesons. $I$ stands for the 
isospin. We denote by a star those states 
listed by the PDG \protect\cite{Pdg04}, and
by a dagger states reported in Refs. \protect\cite{Ani03} and \protect\cite{Bes04}.}
\label{t1}
\begin{center}
\begin{tabular}{c|ccc}
 $I$  & State                 & Mass               & Status \\
\hline
   1  & $a_0(980)$            &  984.7$\pm$1.2     & $*$  \\
      & $a_0(1450)$           &  1474$\pm$19       & $*$  \\
\hline
      & $f_0(600)$            &  400$-$1200        & $*$  \\
      & $f_0(980)$            &  980$\pm$10        & $*$  \\
      & $f_0(1200-1600)$      &  1400$\pm$200      & $\dagger$ \\
      & $f_0(1370)$           &  1200$-$1500       & $*$  \\
   0  & $f_0(1500)$           &  1507$\pm$5        & $*$  \\
      & $f_0(1710)$           &  1714$\pm$5        & $*$  \\
      & $f_0(1790)$ 	      &  1790$^{+40}_{-30}$& $\dagger$ \\
      & $f_0(2020)$           &  1992$\pm$16       & $*$  \\
      & $f_0(2100)$           &  2103$\pm$17       & $*$  \\
      & $f_0(2200)$           &  2197$\pm$17       & $*$  \\
\hline
      & $K_0^*(800)$          &  $\approx$ 800     & $*$  \\
 1/2  & $K_0^*(1430)$         &  1412$\pm$6      & $*$  \\
      & $K_0^*(1950)$         &  1945$\pm$22       & $*$  \\
\end{tabular}
\end{center}
\end{table}

\begin{table}[tbp]
\caption{Mass (QM), in MeV, and flavor dominant component (Flavor) of the 
light scalar mesons considered as $q\bar q$ 
or $qq\bar q \bar q$ states. $I$ stands for the isospin. $nL$ denotes the
radial excitation and the orbital angular momentum corresponding
to the state under consideration.}
\label{t2}
\begin{center}
\begin{tabular}{c|ccc|cc|cc}
  $I$  &\multicolumn{3}{c|}{$q \bar q$} & \multicolumn{2}{c|}{$qq\bar q \bar q$} &  
\multicolumn{2}{c}{Experiment} \\
       & $nL$ & QM   &  Flavor      & QM   & Flavor               &State         & Mass \\
\hline
   1   & $1P$ & 984  & $(n\bar n)$  & 1308 & $(nn\bar n \bar n)$  &$a_0(980)$    & 984.7$\pm$1.2 \\
       & $2P$ & 1587 & $(n\bar n)$  & 1522 & $(ns\bar n \bar s)$  &$a_0(1450)$   & 1474$\pm$19   \\
\hline
       & $1P$ & 413  & $(n\bar n)$  & $-$  & $-$                  &$f_0(600)$    & 400$-$1200   \\
       & $-$  & $-$  & $-$          & 949  & $(nn\bar n \bar n)$  &$f_0(980)$    & 980$\pm$10 \\
       & $1P$ & 1340 & $(s\bar s)$  & $-$  & $-$                  &$f_0(1200-1600)$ & 1400$\pm$200\\
       & $2P$ & 1395 & $(n\bar n)$  & $-$  & $-$                  &$f_0(1370)$   & 1200$-$1500 \\
   0   & $-$  & $-$  & $-$          & 1525 & $(ns\bar n \bar s)$  &$f_0(1500)$   & 1507$\pm$5 \\
       & $3P$ & 1754 & $(n\bar n)$  & $-$  & $-$                  &$f_0(1710)$   & 1714$\pm$5 \\
       & $-$  & $-$  & $-$          & $-$  & $-$                  &$f_0(1790)$   & 1790$^{+40}_{-30}$ \\
       & $2P$ & 1894 & $(s\bar s)$  & $-$  & $-$                  &$f_0(2020)$   & 1992$\pm$16 \\
       & $-$  & $-$  & $-$          & 1915 & $(ss\bar s \bar s)$  &$f_0(2100)$   & 2103$\pm$17 \\
       & $3P$ & 2212 & $(s\bar s)$  & $-$  & $-$                  &$f_0(2200)$   & 2197$\pm$17 \\
\hline
       & $1P$ & 1213 & $(n\bar s)$  & $-$  & $-$                  &$K^*_0(800)$  & $\approx$ 800 \\
   1/2 & $2P$ & 1768 & $(n\bar s)$  & 1295 & $(nn\bar n \bar s)$  &$K^*_0(1430)$ & 1412$\pm$6 \\
       & $3P$ & 2046 & $(n\bar s)$  & 1802 & $(ns\bar s \bar s)$  &$K^*_0(1950)$ & 1945$\pm$22 \\
\end{tabular}
\end{center}
\end{table}

\begin{table}[tbp]
\caption{Mass (QM), in MeV, and flavor dominant component (Flavor) of the 
light isoscalar, isovector and I=1/2  mesons, 
mixing two- and four-quark states as explained in the text.}
\label{t4}
\begin{center}
\begin{tabular}{ccc|ccc|ccc}
\multicolumn{3}{c|}{I=0} & 
\multicolumn{3}{c|}{I=1} & 
\multicolumn{3}{c}{I=1/2}  \\ 
QM & Flavor & Mass &
QM & Flavor & Mass & 
QM & Flavor & Mass \\
\hline
568 &  $(n\bar n)$ & 400$-$1200 &
985 &  $(n\bar n)$ & 984.7$\pm$1.2 &
1113&  $(n\bar s)$ & $\approx$ 800 \\
999 &  $(nn\bar n\bar n)$ & 980$\pm$10 &
$\begin{array}{c}1381\\1530\end{array}$&
$\left. \begin{array}{c} (nn\bar n\bar n)\\ (ns\bar n\bar s) \end{array} \right\}$ &
1474$\pm$19 &
1440&  $(nn\bar n\bar s$) & 1412$\pm$6 \\
1301&  $(s\bar s)$ & 1400$\pm$200 &
1640&  $(n\bar n)$ &      $-$     &
    &              &               \\
1465&  $(n\bar n)$ & 1200$-$1500 &
1868&  $(n\bar n)$ &      $-$    & 
    &              &              \\
1614&  $(ns\bar n\bar s$) & 1507$\pm$5 & 
    &                     &            & 
$\begin{array}{c}1784\\1831\\2060\end{array}$&
$ \left. \begin{array}{c} (n\bar s)\\ (ns\bar s\bar s) \\(n\bar s) 
\end{array} \right\} $ &
1945$\pm$20 \\
$-$ &  $-$         & 1714$\pm$5   &
    &              &              & 
    &              &               \\
1782&  $(n\bar n)$ & 1790$^{+40}_{-30}$ &
    &              &                    & 
    &              &                     \\
1900&  $(s\bar s)$ & 1992$\pm$16 &
    &              &                    & 
    &              &                     \\
1944&  $(ss\bar s \bar s)$ & 2103$\pm$17 &
    &              &                    & 
    &              &                     \\
2224&  $(s\bar s)$ & 2197$\pm$17 &
    &              &                    & 
    &              &                     \\
\end{tabular}
\end{center}
\end{table}

\begin{table}[tbp]
\caption{Flavor wave function components of some of the light scalar mesons.
Masses are given in MeV and the probabilities in \%.}
\label{t5}
\begin{tabular}{|c|ccccc|}
\multicolumn{6}{|c|}{I=0} \\
\hline
QM                       &568      &999       &1301      &1465     &1614   \\
\hline
P($nn\bar n\bar n$)      &15       &82        &$\sim$ 1  &$\sim$1  &$<$1      \\
P($ns\bar n\bar s$)      &$\sim$1  &$\sim$ 1  &23        &$\sim$1  &72  \\
P($n\bar n_{1P}$)        &77       & 9        &14        &$\sim$1  &$<$1   \\
P($s\bar s_{1P}$)        &7        & 7        &58        &4        &22  \\
P($n\bar n_{2P}$)        &$<$1     &$\sim$ 1  &$\sim$ 1  &84       &5   \\
P($s\bar s_{2P}$)        &$<$1     &$<$1      &$<$1      &9        &$<$1   \\
P($ss\bar s\bar s$)      &$<$1     &$<$1      &$\sim$ 1  &$<$1     &$\sim$1   \\
\end{tabular}
\begin{tabular}{|c|ccc|c|cc|}
\multicolumn{4}{|c|}{I=1}  & \multicolumn{3}{c|}{I=1/2}  \\
\hline
QM                  &985      &1381  &1530     &QM                    &1113  &1440 \\
\hline
P($nn\bar n\bar n$) &21       &75    &$\sim$1  &P($nn\bar n\bar s$)   &46   &53 \\
P($ns\bar n\bar s$) &$\sim$1  &4     &90       &P($n\bar s_{1P}$)     &53   &44 \\
P($n\bar n_{1P}$)   &78       &18    &3        &P($n\bar s_{2P}$)     &$<$1 &$<$1 \\
P($n\bar n_{2P}$)   &$<$1     &3     &6        &P($ns\bar s\bar s$)   &$<$1 &$\sim$1 \\
\end{tabular}
\end{table}

\begin{table}[tbp]
\caption{Mass, in MeV, and flavor dominant component of the 
light isoscalar mesons. The mixing with the glueball component is
included as explained in the text.}
\label{t8}
\begin{tabular}{cccc}
State & Exp. & Mass & Flavor \\
\hline
$f_0(600)$            &  400$-$1200        &  568  &  $(n\bar n_{1P})$     \\
$f_0(980)$            &  980$\pm$10        &  999  &  $(nn\bar n\bar n)$   \\
$f_0(1200-1600)$      &  1400$\pm$200      & 1299  &  $(s\bar s_{1P})$     \\
$f_0(1370)$           &  1200$-$1500       & 1406  &  $(n \bar n_{2P})$    \\
$f_0(1500)$           &  1507$\pm$5        & 1611  &  $(ns\bar n\bar s)$   \\
$f_0(1710)$           &  1714$\pm$5        & 1704  &  (glueball)           \\
$f_0(1790)$           &  1790$^{+40}_{-30}$& 1782  &  $(n \bar n_{3P})$    \\
$f_0(2020)$           &  1992$\pm$16       & 1902  &  $(ss \bar s\bar s)$  \\
$f_0(2100)$           &  2103$\pm$17       & 1946  &  $(s \bar s_{2P})$    \\
$f_0(2200)$           &  2197$\pm$17       & 2224  &  $(s \bar s_{3P})$    \\
\end{tabular}
\end{table}

\begin{figure}
\caption{Regge trajectories for the isoscalar mesons. The squares represent
the results of Table \protect\ref{t8}. The lower solid line corresponds to
$n \bar n$ systems and the upper line to $s \bar s$ systems.
The dashed lines correspond to the mass
of those states with a high non$-q\bar q$ component.}
\label{fig1}
\end{figure}

\end{document}